\documentclass[epj]{svjour}
\usepackage{graphics}
\usepackage{epsfig}
\usepackage{psfrag}
\catcode`@=11

\def\@citex[#1]#2{\if@filesw\immediate\write\@auxout{\string\citation{#2}}\fi
  \def\@citea{}\@cite{\@for
\@citeb:=#2\do
    {\@citea\def\@citea{,\penalty\@m}\@ifundefined
      {b@\@citeb}{{\bf ?}\@warning
       {Citation `\@citeb' on page \thepage \space undefined}}%
\hbox{\csname b@\@citeb\endcsname}}}{#1}}

\def\citer{\@ifnextchar [{\@tempswatrue\@citexr}{\@tempswafalse\@citexr[]}}

\def\@citexr[#1]#2{\if@filesw\immediate\write\@auxout{\string\citation{#2}}\fi
  \def\@citea{}\@cite{\@for\@citeb:=#2\do
    {\@citea\def\@citea{--\penalty\@m}\@ifundefined
       {b@\@citeb}{{\bf ?}\@warning
       {Citation `\@citeb' on page \thepage \space undefined}}%
\hbox{\csname b@\@citeb\endcsname}}}{#1}}
\begin{document}
\title{
 CP Violation in B Decays within QCD Factorization\thanks
{Contribution to the International Europhysics Conference on High Energy Physics EPS03, 17-23 July 2003, Aachen, Germany.}}
\author{A. Salim Safir}
\institute{
 Ludwig-Maximilians-Universit\"at M\"unchen, Sektion Physik,\\
Theresienstra\ss e 37, D-80333 Munich, Germany}

\date{{\tt LMU 26/03}}

\abstract{  
We analyze the extraction of weak phases from CP violation in 
$B\to\pi^+\pi^-$ decays. By combining the information on mixing induced CP violation in $B\to\pi^+\pi^-$, namely $S_{\pi\pi}$, with the precision observable $\sin 2\beta$ obtained from the ``gold-plated''mode $B\to\psi K_S$, we propose the determination of the unitarity triangle.
We also discuss 
alternative ways to analyze $S_{\pi\pi}$ which can be useful if new physics 
affects $B_d$--$\bar B_d$ mixing. Predictions and uncertainties for $r$ and 
$\phi$ in QCD factorization are examined in detail. It is pointed out that a 
simultaneous expansion in $1/m_b$ and $1/N_C$ leads to interesting 
simplifications. At first order infrared divergences are absent, while the 
most important effects are retained. Independent experimental tests of the 
factorization framework are briefly discussed.
}


\authorrunning{A. Salim Safir}
\titlerunning{
CP Violation in B Decays within QCD Factorization
}

\maketitle

\section{Introduction}
\label{Intro}
The main goal of the current experimental program at the SLAC
and KEK $B$-meson factories is a stringent test of the standard model 
description of CP violation. In the future this aim will be pursued
with higher precision measurements from hadron machines at
Fermilab and CERN. 

A crucial benchmark is the time-dependent CP violation
in $B\to\psi K_S$ decays, which allows us to infer the CKM phase $\beta$
with negligible hadronic uncertainties.
Likewise of central importance for obtaining additional information on
CKM parameters is the time-dependent CP violation,
both mixing-induced ($S_{\pi\pi}$) and direct ($C_{\pi\pi}$), in $B\to\pi^+\pi^-$.
However, in this case the extraction of weak phases is complicated
by the so-called penguin pollution, leading to a hadronic model dependent estimate of the CP asymmetries in $B\to\pi^+\pi^-$.
A possible strategy to circumvent this problem is the use of symmetry argument 
\citer{GL,GR},
such as the isospin or the SU(3) symmetry. 
However these methods are actually very limited which is likely to prevent a successful realization.

In this talk, we present the result of \cite{safir}, where a new way of exploring informations from the mixing induced CP violation parameter $S_{\pi\pi}$ in the $B\to\pi^+\pi^-$ mode, combined with the precision  observable $sin 2\beta$ was suggested in order to explore the CKM unitarity triangle. Our estimate of the penguin parameters are carried out in QCD factorization and confronted to other independent estimates.

\section{Basic Formulas}
The time-dependent CP asymmetry in $B\to\pi^+\pi^-$ decays
is defined by
\begin{eqnarray}\label{acppipi}
A^{\pi\pi}_{CP}(t) &=& 
\frac{B(B(t)\to\pi^+\pi^-)-B(\bar B(t)\to\pi^+\pi^-)}{
  B(B(t)\to\pi^+\pi^-)+B(\bar B(t)\to\pi^+\pi^-)}, \nonumber \\
&=& - S_{\pi\pi}\, \sin(\Delta m_B t) + C_{\pi\pi}\, \cos(\Delta m_B t),
\end{eqnarray}
where
\begin{equation}\label{scxi}
S_{\pi\pi}=\frac{2\, {\rm Im}\xi}{1+|\xi|^2},\,\,\,
C_{\pi\pi}=\frac{1-|\xi|^2}{1+|\xi|^2},\,\,\,
\xi=e^{-2 i\beta}\,\frac{e^{-i\gamma}+P/T}{e^{+i\gamma}+P/T}.
\end{equation}
In terms of the Wolfenstein parameters $\bar\rho$ and $\bar\eta$
the CKM phase factors read
\begin{equation}\label{gambetre}
e^{\pm i\gamma}=\frac{\bar\rho\pm i \bar\eta}{\sqrt{\bar\rho^2+\bar\eta^2}},
\,\,\,
e^{-2 i\beta}=\frac{(1-\bar\rho)^2 -\bar\eta^2 - 2 i\bar\eta(1-\bar\rho)}{
                     (1-\bar\rho)^2 + \bar\eta^2}.
\end{equation}
The penguin-to-tree ratio $P/T$ can be written as
$P/T=r e^{i\phi}/\sqrt{\bar\rho^2+\bar\eta^2}$.
The real parameters $r$ and $\phi$ defined in this way are
pure strong interaction quantities without further dependence
on CKM variables.

For any given values of $r$ and $\phi$ a measurement of $S_{\pi\pi}$ 
defines a curve in the ($\bar\rho$, $\bar\eta$)-plane.
Using the relations above this constraint is given by the
equation
\begin{equation}\label{srhoeta}
S_{\pi\pi}=\frac{2\bar\eta [\bar\rho^2+\bar\eta^2-r^2-\bar\rho(1-r^2)+
       (\bar\rho^2 +\bar\eta^2-1)r \cos\phi]}{((1-\bar\rho)^2+\bar\eta^2)
         (\bar\rho^2+\bar\eta^2+r^2 +2 r\bar\rho \cos\phi)}
\end{equation}
Similarly the relation between $C_{\pi\pi}$ and $\bar\rho$, $\bar\eta$ is straightforward.
The current experimental results for $S_{\pi\pi}$ and $C_{\pi\pi}$ are
\begin{eqnarray}\nonumber
\begin{array}{rcc}
           &({\rm BaBar} \cite{BABAR1})  & ({\rm Belle} \cite{BELLE1}) \\
S_{\pi\pi} = & +0.02\pm 0.34\pm 0.05  &-1.23\pm 0.41 ^{+0.08}_{-0.07}   \\
C_{\pi\pi} = & -0.30\pm 0.25\pm 0.04  &\,\,\,\,\, -0.77\pm 0.27\pm 0.08     
\end{array}
\end{eqnarray}
A recent preliminary update from BaBar \cite{HFAG} gives
$S_{\pi\pi}=-0.40\pm 0.22\pm 0.03$, and $C_{\pi\pi}=-0.19\pm 0.19\pm 0.05$.

The penguin parameter $r\, e^{i\phi}$ has been computed
in \cite{BBNS1} in the framework of QCD factorization.
The result can be expressed in the form
\begin{equation}\label{rqcd}
r\, e^{i\phi}= -
\frac{a^c_4 + r^\pi_\chi a^c_6 + r_A[b_3+2 b_4]}{
 a_1+a^u_4 + r^\pi_\chi a^u_6 + r_A[b_1+b_3+2 b_4]},
\end{equation}
where we neglected the very small effects from electroweak
penguin operators. A recent analysis gives \cite{safir,BS}
\begin{equation}\label{rphi}
r=0.107\pm 0.031, \qquad \phi=0.15\pm 0.25,
\end{equation}
where the error includes an estimate of potentially important
power corrections.
In order to obtain additional insight into the structure of
hadronic $B$-decay amplitudes, it will be also interesting to extract these 
quantities from other B-channels, or using other methods.
In this perspective, we have considered them in a simultaneous expansion 
in $1/m_b$ and $1/N_C$ ($N_C$ is the number of colours).

As stated above, the most important contributions in ($r,\phi$) are the factorization coefficients $a_{1,4,6}$ and the weak annihilations ones $b_{1,3,4}$ as shown\footnote{
both corrections
depend on the unknown power corrections effects, described by phenomenological quantities, $X_{H,A}=\left(1+\rho_{H,A}\,e^{i\phi_{H,A}}\right)\ln\frac{m_B}{\Lambda_h}$.} in (\ref{rqcd}).
Expanding these coefficients to first order in $1/m_b$ and
$1/N_C$ we find
\begin{eqnarray}\label{a1mn}
a_1\,&\dot =&\, C_1 +\frac{C_2}{N_C}\left[1+\frac{C_F\alpha_s}{4\pi}V_\pi\right]
           +\frac{C_2}{N_C}\frac{C_F\pi\alpha_s}{N_C}H_{\pi\pi,2},
\\
a^p_4\,&\dot=&\, C_4 +\frac{C_F\alpha_s}{4\pi} \frac{P^p_{\pi,2}}{N_C},
\qquad
r^\pi_\chi a^p_6\,\dot=\, r^\pi_\chi C_6,
\qquad 
b_{1,3,4}\,\dot=\,0,\nonumber
\end{eqnarray}
where $H_{\pi\pi,2}$ is the leading-twist effect in the hard spectator scattering. We observe that to this order in the double expansion, the uncalculable power  $H_{\pi\pi,3}(\sim X_H)$ does not appear in $a_1$, to
which it only contributes at order $1/m_b N_C$. Using our default input parameters, one obtaines the central value \cite{safir}: 
$(r_{N_C},\phi_{N_C})=(0.084,0.065)$,
which seems to be in a good agreement with the standard QCD factorization framework at the next-to-leading order.

As a second cross-check, one can extract
$r$ and $\phi$ from $B^+\to\pi^+\pi^0$ and  $B^+\to\pi^+ K^0$, leading to the central value \cite{safir} $(r_{SU3},\phi_{SU3})=(0.081, 0.17)$, in agreement with the above results\footnote{one can compare also the $r_{SU3}$ to its experimental value $r_{SU3}^{exp}=0.099\pm0.014$.}, although their definitions differ slightly from $(r,\phi)$ (see \cite{safir} for further discussions). 
\section{Exploring the Unitarity Triangle in the SM and beyond}
In this section we discuss the determination of the unitarity
triangle by combining the information from $S_{\pi\pi}$
with the value of $\sin 2\beta$, well known from the ``gold-plated'' mode 
$B\to J/\Psi K_S$.
\begin{figure}[t!]
\psfrag{S}{$S$}
\psfrag{etabar}{\hspace*{0.5cm}$\bar\eta$}
\begin{center}
\epsfig{file=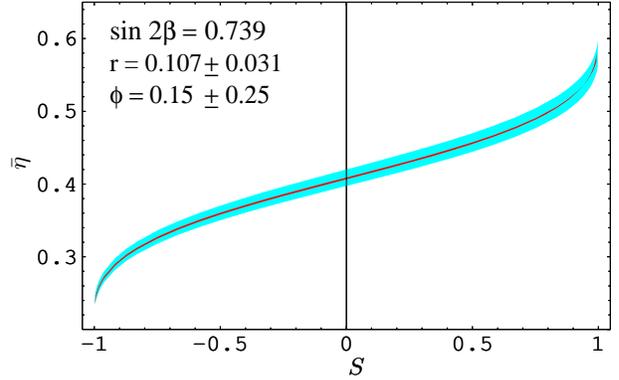,width=8cm,height=5cm}
\end{center}
{\caption{\it CKM phase $\bar\eta$ as a function of 
$S_{\pi\pi}$.
The dark (light) band reflects the theoretical uncertainty
in the parameter  $\phi$ ($r$).      
\label{fig:etabspp}}}
\end{figure}
The angle $\beta$ of the unitarity triangle is given by
\begin{equation}\label{taus2b}
\tau\equiv\cot\beta=\sin 2 \beta\Bigg(1-\sqrt{1-\sin^2 2\beta}\Bigg)^{-1}.
\end{equation}
The current world average \cite{HFAG} 
\begin{equation}\label{sin2bexp}
\sin 2\beta =0.739\pm 0.048,
\end{equation}
implies
$\tau=2.26\pm 0.22.$
Given a value of $\tau$, $\bar\rho$ is related to $\bar\eta$
by
$\bar\rho = 1-\tau\, \bar\eta$.
The parameter $\bar\rho$ may thus be eliminated from $S_{\pi\pi}$
in (\ref{srhoeta}), which can be solved for $\bar\eta$ to yield
\begin{eqnarray}\label{etataus}
\bar\eta &&=\frac{1}{(1+\tau^2)S_{\pi\pi}} 
\Bigg[(1+\tau S_{\pi\pi})(1+r \cos\phi) \\
&&
\hspace*{-0.5cm}-\sqrt{(1-S_{\pi\pi}^2)(1+r^2+2 r\cos\phi)-
  (1+\tau S_{\pi\pi})^2 r^2 \sin^2\phi}\Bigg].\nonumber 
\end{eqnarray}
The two observables $\tau$ (or $\sin 2\beta$) and $S_{\pi\pi}$ determine
$\bar\eta$ and $\bar\rho$ once the theoretical penguin parameters
$r$ and $\phi$ are provided. It is at this point that some theoretical
input is necessary. We will now consider the impact of the
parameters $r$ and $\phi$, and of their uncertainties, on the analysis.

We first would like to point out that the sensitivity of $\bar\eta$
in (\ref{etataus}) on the strong phase $\phi$  is rather mild. In fact,
the dependence on $\phi$ enters in (\ref{etataus}) only at second order, and hence suppressed for small $\phi$. This nice feature is {\it la bienvenue} because the estimate of the strong phase is difficult.

The determination of $\bar\eta$ as a function of $S_{\pi\pi}$ is shown
in Fig. \ref{fig:etabspp},
which displays the theoretical uncertainty from the penguin
parameters $r$ and $\phi$ in QCD factorization.

In the determination of $\bar\eta$ and $\bar\rho$ described
here discrete ambiguities do in principle arise, however they are ruled out using the standard fit of the unitarity triangle (see \cite{safir} for further discussions).

Up to now, our analysis was carried out within the standard model. However, in the presence of new physics this may no longer be valid.
To be specific we shall assume the plausible scenario where the
new physics contributions modify the phase of $B_d$-$\bar B_d$ mixing $\phi_d$,
whereas the $B$ decay amplitudes remain unchanged.
The CP asymmetry in $B\to J/\Psi K_S$ (\ref{sin2bexp}) must then 
be interpreted as the quantity $\sin 2\phi_d$.
Since we can no longer relate $\sin 2\phi_d$ to $\bar\rho$ and $\bar\eta$,
we should fix it to the experimental value in (\ref{sin2bexp})
when using (\ref{scxi}), where $\beta$ is to be replaced by $\phi_d$.
A similar analysis has already been carried out in \cite{FIM}.

Writing $(\bar\rho,\bar\eta)=R_b (\cos\gamma, \sin\gamma)$ 
and
$R_b\equiv
\sqrt{\bar\rho^2+\bar\eta^2}
$,
we can express $S_{\pi\pi}$ as 
\begin{equation}\label{srbgam}
S_{\pi\pi}={\rm Im}\left[e^{-2i\phi_d}
\frac{(R_b\cos\gamma+r\cos\phi-i R_b\sin\gamma)^2+\kappa^2}{
(R_b\cos\gamma+r\cos\phi)^2 +R^2_b\sin^2\gamma +\kappa^2 }\right],
\end{equation}
where $\kappa= r\,sin\phi$. From this relation, for given values of $r$ and $\phi$,
and using the experimental results for $S_{\pi\pi}$, $\sin 2\phi_d$ and $R_b$,
$\gamma$  can be determined and hence $\bar\rho$ and $\bar\eta$.

\begin{figure}[t]
\psfrag{rob}{$\bar\rho$}
\psfrag{etab}{\hspace*{0.2cm}$\bar\eta$}
\begin{center}
\epsfig{file=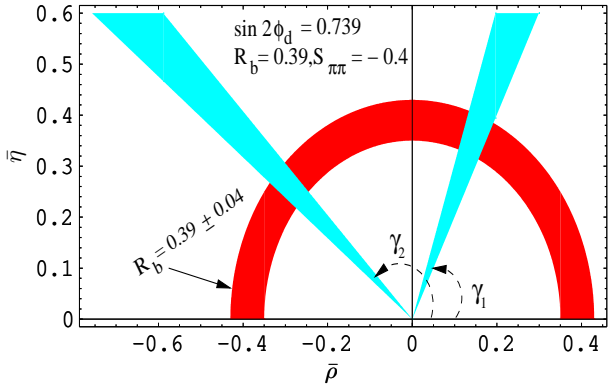,width=8cm,height=5cm}
\end{center}
{\caption{\it Determination of the unitarity triangle from
$S_{\pi\pi}$, $R_b$, $\sin 2\phi_d$,and the penguin parameters
as in Fig.\ref{fig:etabspp}.
The resulting ranges of the CKM angle $\gamma$
are $\gamma_1=67^\circ \pm 4^\circ$ 
and $\gamma_2=138^\circ \pm 4^\circ$.
The light bands reflect the uncertainty in $r$ and $\phi$.
\label{fig:sppnp}}}
\end{figure}

Experimentally one has \cite{HFAG,CKMF}
$\sin 2\phi_d=0.739\pm 0.048$ and $R_b=0.39\pm 0.04$.
As emphasized in \cite{FIM}, there is a discrete ambiguity
in the sign of $\cos 2\phi_d$, which yields two different
solutions for $\gamma$. The larger value of $\gamma$ will be 
obtained for negative $\cos 2\phi_d$.
The analysis is represented in Fig. \ref{fig:sppnp}, assuming
a particular scenario for illustration and displaying the
impact of the theoretical uncertainty in $r$ and $\phi$.

\section{Summary}
In this talk, we have proposed strategies to extract
information on weak phases from CP violation
observables in $B\to\pi^+\pi^-$ decays even in the presence
of hadronic contributions related to penguin amplitudes.
Assuming knowledge of the penguin pollution, an efficient use of mixing-induced CP violation in $B\to\pi^+\pi^-$
decays, measured by $S_{\pi\pi}$, can be made by combining it with the corresponding
observable from $B\to\psi K_S$, $\sin 2\beta$, to obtain the unitarity triangle parameters $\bar\rho$ and $\bar\eta$.
The sensitivity on the hadronic quantities, which have typical
values $r\approx 0.1$, $\phi\approx 0.2$, is very weak.
In particular, there are no first-order corrections in $\phi$.
For moderate values of $\phi$ its effect is negligible.
An alternative analysis of $S_{\pi\pi}$ especially suitable for the presence 
of a new-physics phase in $B_d$--$\bar B_d$ mixing was discussed.

Concerning our penguin 
parameters, 
namely $r$ and $\phi$, they were 
investigated systematically within the QCD factorization framework.
To validate our theoretical predictions, we have calculate these parameters in the $1/m_b$ and $1/N_C$ expansion, which exhibits a good framework to control the uncalculable power corrections, namely $X_{H,A}$, in the factorization formalism. As an alternative proposition,  we have also considered to extract $r$ and $\phi$ from other $B$ decay channels, such as $B^+\to\pi^+\pi^0$ and $B^+\to\pi^+ K^0$, relying on the SU(3) argument. Using these three different approaches, we found a compatible picture in estimating these hadronic parameters.

Finally, we hope that in the forthcoming experimental measurements of CP violation in $B\to\pi^+\pi^-$ decays, our analysis will be useful in achieving 
a reliable control over these unknown penguin parameters.

\subsection*{Acknowledgments}
I would like to thank G. Buchalla for the fruitful collaboration. This work is supported by the Deutsche
Forschungsgemeinschaft (DFG) under contract BU 1391/1-2.


\end{document}